\documentclass[reprint,floatfix,twocolumn,superscriptaddress,showpacs, aps, prx]{revtex4-2} 

\usepackage{graphicx} 
\usepackage{natbib} 
\usepackage[usenames,dvipsnames]{color} 
\usepackage{amsmath} 

\usepackage[urlcolor=blue, hyperindex, colorlinks, bookmarks=true,linkcolor=black,citecolor=black]{hyperref} 
\usepackage{dcolumn}
\usepackage{amssymb} 
\usepackage{soul} 
\usepackage{ifthen} 
\usepackage{bbm}
\usepackage{comment}
\usepackage[ampersand]{easylist}
\usepackage{todonotes}
\usepackage{upgreek}

\newcommand{\bra}[1]{\langle{#1}|}
\newcommand{\ket}[1]{|{#1}\rangle}

\def\l{\left}
\def\r{\right}

\def\be#1\ee{\begin{equation}#1\end{equation}}
\def\ba#1\ea{\begin{align}#1\end{align}}
\def\bg#1\eg{\begin{gather}#1\end{gather}}
\def\t{\text}

\def\shownoteal{1} 
\newcommand{\nal}[1]{\ifthenelse{\shownoteal=1}{\textcolor{red}{[[#1]]}}{}}
\def\shownoteay{0} 
\newcommand{\nay}[1]{\ifthenelse{\shownoteay=1}{\textcolor{orange}{[[#1]]}}{}}
\def\showaddmat{1} 
\newcommand{\addmat}[1]{\ifthenelse{\showaddmat=1}{\textcolor{Gray}{[[#1]]}}{}}
\def\shownote{1} 
\newcommand{\note}[1]{\ifthenelse{\shownote=1}{\textcolor{Red}{[[#1]]}}{}}

\begin{document}

\title{Decoherence of a tunable capacitively shunted flux qubit}

\author{Robbyn Trappen}
\thanks{These two authors contributed equally. rtrappen@uwaterloo.ca, x35dai@uwaterloo.ca}
\affiliation{Institute for Quantum Computing, and Department of Physics and Astronomy,
University of Waterloo, Waterloo, ON, Canada N2L 3G1}
\author{Xi Dai}
\thanks{These two authors contributed equally. rtrappen@uwaterloo.ca, x35dai@uwaterloo.ca}
\affiliation{Institute for Quantum Computing, and Department of Physics and Astronomy, University of Waterloo, Waterloo, ON, Canada N2L 3G1}
\author{M. Ali Yurtalan}
\affiliation{Institute for Quantum Computing, and Department of Physics and Astronomy, University of Waterloo, Waterloo, ON, Canada N2L 3G1}
\affiliation{Department of Electrical and Computer Engineering,
University of Waterloo, Waterloo, ON, Canada N2L 3G1}
\author{Denis Melanson}
\affiliation{Institute for Quantum Computing, and Department of Physics and Astronomy, University of Waterloo, Waterloo, ON, Canada N2L 3G1}
\author{Daniel M. Tennant}
\author{Antonio J. Martinez}
\affiliation{Institute for Quantum Computing, and Department of Physics and Astronomy, University of Waterloo, Waterloo, ON, Canada N2L 3G1}
\author{Yongchao Tang}
\affiliation{Institute for Quantum Computing, and Department of Physics and Astronomy, University of Waterloo, Waterloo, ON, Canada N2L 3G1}
\author{Joseph Gibson}
\affiliation{Department of Physics and Astronomy, Dartmouth College, Hanover, 03755, USA}
\author{Jeffrey A. Grover}
\affiliation{Research Laboratory of Electronics, Massachusetts Institute of Technology, Cambridge, Massachusetts 02139, USA}
\author{Steven M. Disseler}
\author{James I. Basham}
\author{Rabindra Das}
\author{David K. Kim}
\author{Alexander J. Melville}
\author{Bethany M. Niedzielski}
\author{Cyrus F. Hirjibehedin}
\author{Kyle Serniak}
\author{Steven J. Weber}
\author{Jonilyn L. Yoder}
\affiliation{Lincoln Laboratory, Massachusetts Institute of Technology, Lexington, Massachusetts, 02421, USA}
\author{William D. Oliver}
\affiliation{Research Laboratory of Electronics, Massachusetts Institute of Technology, Cambridge, Massachusetts 02139, USA}
\affiliation{Lincoln Laboratory, Massachusetts Institute of Technology, Lexington, Massachusetts, 02421, USA}
\author{Daniel A. Lidar}
\affiliation{Departments of Electrical \& Computer Engineering, Chemistry, and Physics, and Center for Quantum Information Science \& Technology, University of Southern California, Los Angeles, California 90089, USA}
\author{Adrian Lupascu}\thanks{adrian.lupascu@uwaterloo.ca}
\affiliation{Institute for Quantum Computing, and Department of Physics and Astronomy,
University of Waterloo, Waterloo, ON, Canada N2L 3G1}
\affiliation{Waterloo Institute for Nanotechnology, University of Waterloo, Waterloo, ON, Canada N2L 3G1}

\date{ \today}

\begin{abstract}
Quantum annealing is a method to solve optimization problems that leverages quantum tunneling in a coupled qubit system. We present a detailed study of the coherence of a tunable capacitively-shunted flux qubit, designed for coherent quantum annealing applications. We find that for high qubit frequencies, thermal noise in the bias line makes a significant contribution to the relaxation, arising from the design choice to experimentally explore both fast annealing and high-frequency control. The measured dephasing rate is primarily due to intrinsic low-frequency flux noise in the two qubit loops, with additional contribution from the low-frequency noise of control electronics used for fast annealing. Our results characterize decoherence in a realistic setup for quantum annealing and are relevant for ongoing efforts toward building superconducting quantum annealers with increased coherence.

\end{abstract}

\maketitle

\section*{Introduction}
Quantum annealing (QA)~\cite{kadowaki_1998_quantumannealingtransverse,farhiQuantumAdiabaticEvolution2001,albash_2018_adiabaticquantumcomputation,haukePerspectivesQuantumAnnealing2020a} is a computational paradigm that shows promise for outperforming classical computers in solving certain computational tasks, such as optimization, machine learning~\cite{amin_2018_quantumboltzmannmachine} and quantum simulation~\cite{king_2021_scalingadvantagepathintegral}. In comparison to the more commonly-pursued gate-model quantum computation (GMQC), QA is in principle computationally equivalent~\cite{mizel_2007_simpleproofequivalence,aharonov_2008_adiabaticquantumcomputation}, and is more amenable to scaling up in the near term, since it does not require individual dynamic control of each qubit. Among others~\cite{islamEmergenceFrustrationMagnetism2013,goto_2016_bifurcationbasedadiabaticquantum,ebadiQuantumOptimizationMaximum2022,puri_2017_quantumannealingalltoall}, one of the most prominent implementations of quantum annealing is based on superconducting flux qubits~\cite{harrisExperimentalDemonstrationRobust2010,johnson_2011_quantumannealingmanufactured,quintanaSuperconductingFluxQubits2017,novikovExploringMoreCoherentQuantum2018}. Benefiting from the compatibility of superconducting qubit with industry-scale fabrication and the relatively low control requirements (with only two global controls used to adjust the amplitude of the transverse and Ising terms), commercial quantum annealers have been made available by DWave, with the latest generation device containing more than 5,000 qubits~\cite{QPUSpecificCharacteristicsDWave}. 

One of the open questions on quantum annealing is regarding the role of noise and coherence times~\cite{amin_2009_rolesinglequbitdecoherence,albash_2018_adiabaticquantumcomputation}. Depending on the coupling strength and the instantaneous state of the annealer, the same noise source could cause different effects \cite{ashhab_2006_decoherencescalableadiabatic}. If the noise is weakly coupled, the annealer primarily suffers from thermal relaxation and excitation for most forms of noise, while if the noise is strongly coupled, thermal relaxation and excitation could either improve~\cite{dickson_2013_thermallyassistedquantum} or hamper annealing depending on the exact problem~\cite{mishra_2018_finitetemperaturequantum}. There is recently increasing experimental evidence of strong coupling being relevant for quantum annealing, especially when the system spectral gap becomes small~\cite{boixo_2016_computationalmultiqubittunnelling,bando_2022_breakdownweakcouplinglimit}. While some previous research suggests that strong coupling to the environment renders coherent quantum annealing impossible~\cite{albash_decoherence_2015}, other work points out that incoherent quantum-assisted tunneling is still possible, which could still lead to a quantum advantage~\cite{denchev_2016_whatcomputationalvalue}. To further understand the range of applicability of different noise models, it is important to understand the underlying sources of the noises in the device. 

There have been many previous studies that contributed to the understanding of noise in superconducting flux qubits and superconducting circuits in general~\cite{siddiqiEngineeringHighcoherenceSuperconducting2021}. Various noise sources have been investigated in detail, including intrinsic $1/f$ flux noise~\cite{yoshihara_2006_decoherencefluxqubits,kakuyanagiDephasingSuperconductingFlux2007,bialczak_2007_fluxnoisejosephson,lanting_2009_geometricaldependencelowfrequency,yoshihara_2014_fluxqubitnoise,quintana_2017_observationclassicalquantumcrossover,braumuller_2020_characterizingoptimizingqubit}, photon noise in the readout resonator~\cite{bertetDephasingSuperconductingQubit2005,schusterAcStarkShift2005,rigettiSuperconductingQubitWaveguide2012,yan_2016_fluxqubitrevisited,yanDistinguishingCoherentThermal2018}, dielectric loss~\cite{martinisDecoherenceJosephsonQubits2005,wang_2015_surfaceparticipationdielectric,gambettaInvestigatingSurfaceLoss2017,somoroff_2021_millisecondcoherencesuperconducting,changTunableSuperconductingFlux2022,changReproducibilityControlSuperconducting2022,sunCharacterizationLossMechanisms2023} and junction critical-current noise~\cite{vanharlingenDecoherenceJosephsonjunctionQubits2004,yan_2012_spectroscopylowfrequencynoise}. While decoherence due to extrinsic flux noise was identified as a possible limitation in early theoretical work \cite{van_der_wal_engineering_2003}, this source is effectively managed in modern experiments with gate-based devices. Building on this understanding, improved designs of the flux qubits have been developed, including the persistent-current qubit ~\cite{mooij_1999_josephsonpersistentcurrentqubit,orlando_1999_superconductingpersistentcurrentqubit,vanderwal_2000_quantumsuperpositionmacroscopic}, the capacitively-shunted flux qubit (CSFQ)\cite{you_2007_lowdecoherencefluxqubit,steffen_2010_highcoherencehybridsuperconducting,yan_2016_fluxqubitrevisited,yurtalan_2021_characterizationmultileveldynamics}, and fluxonium~\cite{manucharyan_2009_fluxoniumsinglecooperpair,nguyen_2019_highcoherencefluxoniumqubit,somoroff_2021_millisecondcoherencesuperconducting}, which brought the coherence times of flux qubits to the milliseconds timescale. 

Commercial annealers use one of the earliest variants of flux qubit, the rf-SQUID qubit~\cite{leggett_1987_dynamicsdissipativetwostate,friedman_2000_quantumsuperpositiondistinct}. The coherence time of these qubits are limited to about $15$ nanoseconds~\cite{ozfidan_2020_demonstrationnonstoquastichamiltonian}, due to a number of reasons, including requirements to simplify control calibration and to allow large coupling strength for annealing applications~\cite{harrisExperimentalDemonstrationRobust2010}. Combined with the low control bandwidth of commercial annealers, this limits the capability for detailed coherence studies. Recently the CSFQ has gained prominence as a suitable candidate that features improved coherence time while having strong enough coupling strength required for annealing~\cite{yan_2016_fluxqubitrevisited,weber_2017_coherentcoupledqubits,novikovExploringMoreCoherentQuantum2018,groverFastLifetimePreservingReadout2020a,khezriAnnealpathCorrectionFlux2021,khezri_2022_customizedquantumannealing}. However, previous studies of coherence in CSFQ use designs without the full flux tunability and controllability required for annealing~\cite{yan_2016_fluxqubitrevisited, weber_2017_coherentcoupledqubits,chang_2023_tunablesuperconductingflux}, thus do not expose the qubits to all the sources of noise possible in a realistic quantum annealing device. 

In this work, we study the coherence of a single tunable capacitively-shunted flux qubit (CSFQ), designed to be incorporated into a large-scale annealer. Whereas devices in previous studies consisted of a single patterned layer, our device utilizes a two-layer fabcrication process with the qubit chip on one layer and the readout and control circuitry (further discussed in the "The CSFQ Device" subsection of Results and Discussion). We measure the coherence times of the CSFQ for a wide range of flux biases, and model the results with a comprehensive list of noise sources. Although the measured coherence times only apply in the weak-coupling limits, the inferred noise power provides a solid basis for future coherence studies in more complex settings. We found that relaxation is dominated by $1/f$ flux noise up to about $\sim3~\text{GHz}$, with additional contributions from bias line thermal noise and possibly two-level system (TLS) defects~\cite{mullerUnderstandingTwolevelsystemsAmorphous2019}, while dephasing is dominated by intrinsic flux noise. Additionally, we note that electronics required for high bandwidth control introduces challenges related to dephasing~\cite{ithier_2005_decoherencesuperconductingquantum}. Our results are immediately relevant to upcoming experiments that explore coherent QA based on CSFQs, and could inform the design of other variants of flux qubits, such as fluxonium. 

\section*{Results and Discussion}
\subsection*{The CSFQ device}\label{sec:Device}
\begin{figure}
    \centering
    \includegraphics[width=3.4in]{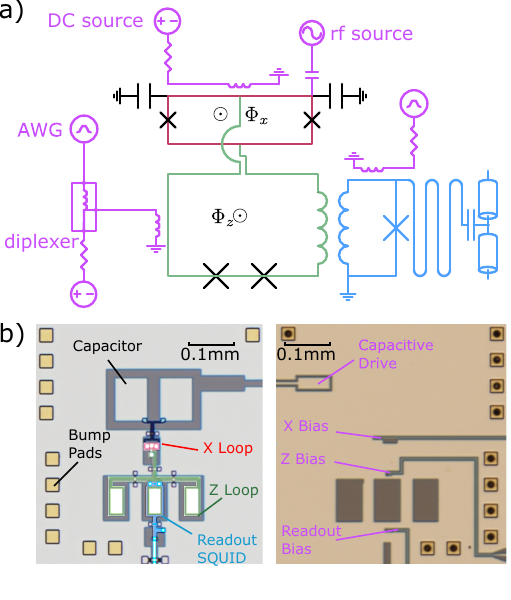}
    \caption{The capacatively shunted flux qubit device. (a)~Schematic of the CSFQ, with the $X$-loop in red, $Z$-loop in green, readout circuit in blue and control circuit in purple. Readout is performed by measuring the transmission through an rf-SQUID coupled inductively to the qubit.(b)~Optical images of the qubit and interposer chips, around the CSFQ. The qubit and readout-SQUID loops are highlighted in false color. The two chips are flip-chip bonded with Indium bumps.}
    \label{fig:CircuitChip}
\end{figure}

The measured CSFQ, together with the control and readout circuitry is schematically shown in Fig.~\ref{fig:CircuitChip}(a). The CSFQ has a main loop and a secondary split-junction loop. In the two-state approximation, the CSFQ has the Hamiltonian
\begin{align}
    H_\t{q} &=-\frac{\hbar\Delta(\Phi_x)}{2}\sigma_x-I_p(\Phi_x)[\Phi_z-\Phi_z^\t{sym}(\Phi_x)]\sigma_z,\label{eq:TwoLevelHam}
\end{align}
where $\hbar$ is the reduced Planck's constant, $I_p$ is the persistent current in the qubit main loop, $\Delta$ is the tunneling amplitude between the two persistent current states, $\Phi_z^\t{sym}$ is the effective flux bias symmetry point, and $\sigma_{z,x}$ are the qubit Pauli operators in the persistent current basis. The flux bias $\Phi_x$ is the externally applied flux in the secondary loop. The flux bias $\Phi_z$ is the effective flux bias in the main loop, defined such that the native crosstalk from the $X$-loop $\Phi_x/2$ is accounted for~\cite{orlando_1999_superconductingpersistentcurrentqubit}. At the symmetry point $\Phi_z=\Phi_z^\t{sym}$, the qubit eigenstates are symmetric and anti-symmetric combinations of the persistent-current basis states corresponding to current flowing in opposite directions in the $Z$-loop. To a large extent $\Phi_z$ controls the longitudinal field strength while $\Phi_x$ controls the transverse field strength, hence the name $Z$ and $X$ respectively. The $X$-loop has a symmetrized design, allowing nearly independent control of the qubit's transverse and longitudinal field (see Supplementary Note 1).

The external flux biases are controlled by currents through on-chip flux-bias lines, which are supplied by DC voltage sources and arbitrary waveform generators (AWGs) at room temperature, combined through cryogenic diplexers. In the experiment reported here, the high-frequency port of the diplexer that controls the $X$-loop was unresponsive, and is hence neglected in the schematic in Fig.~\ref{fig:CircuitChip}(a). In addition to flux biasing, the CSFQ is capacitively coupled to an rf source, allowing microwave excitation of the qubit. 

Readout of the qubit is achieved by coupling the qubit $Z$-loop inductively to an rf-SQUID-terminated $\lambda/4$-resonator, coupled to a transmission line. The external flux bias of the rf-SQUID, $\Phi_r$ is controlled by current through the on-chip bias line. When $\Phi_r=0$, the resonator is at its maximum frequency and the qubit-resonator coupling is linear. At this point, the resonator experiences a qubit-state-dependent dispersive shift. Away from zero bias, the resonator frequency has an increased flux-sensitivity, which can be used for persistent current basis readout of the qubit. 

The device is fabricated at MIT Lincoln Laboratory using a flip-chip process~\cite{rosenberg_2017_3dintegratedsuperconducting}, combining a high-coherence qubit chip hosting the CSFQ, and a control chip that hosts the readout and control circuitry. Optical images of the two chips are shown in Fig.~\ref{fig:CircuitChip}(b). 

\subsection*{Device characterization}\label{sec:Characterization}
In this subsection, we discuss the characterization measurements for the CSFQ. We note that details of the device characterization have been presented in a related manuscript~\cite{dai_dissipative_2022}, and we focus the discussion here on aspects that concern annealing implementation with the CSFQ.

\begin{figure}
    \centering
    \includegraphics[width=3.4in]{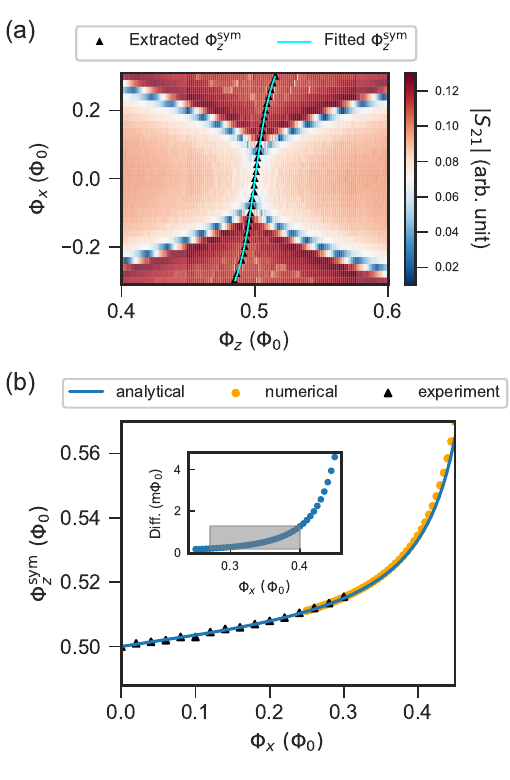}
    \caption{Characterization of the symmetry point as a function of flux bias. (a)~Transmission measurement versus the qubit external flux biases $\Phi_z, \Phi_x$. The black markers correspond to extracted $\Phi_z^\t{sym}$ by finding the point of reflection symmetry for each $\Phi_x$ trace in the transmission measurement, and the blue line is a fit to the analytical expression. (b) $Z$-loop bias symmetry point $\Phi_z^\t{sym}$ as a function of $\Phi_x$, extracted from the measurement shown in panel (a) (black triangles), and calculated from the analytical (blue line) and numerical (orange dots) model respectively. The inset shows the difference between the analytical and numerical models, and the gray shaded region corresponds to the range of $\Phi_x$ in which coherence measurements are performed in this work. The difference in $Z$-loop symmetry point from the analytical and numerical models is always below $1.5$ m$\Phi_0$ in this range, and approaches zero as $\Phi_x$ is reduced towards zero. }
    \label{fig:Calibration}
\end{figure}

The experiments were performed with the device cooled down in a dilution refrigerator, with base temperature of about 10 mK (see Supplementary Note 2 for a detailed diagram of the setup). The crosstalk between the flux-bias lines was measured using the iterative procedure introduced in Ref.~\cite{dai_2021_calibrationfluxcrosstalk}. After three iterations, the error in flux biasing is expected to be about $1~\t{m}\Phi_0/\Phi_0$, with $\Phi_0$ being the flux quantum. In Fig.~\ref{fig:Calibration} we show the transmission through the line coupled to the resonator, at a frequency near the resonator maximum frequency versus the calibrated external fluxes $\Phi_z, \Phi_x$. This measurement shows that the $\Phi_z$ symmetry point has a non-linear dependence on $\Phi_x$. This is due to the asymmetry between the two junctions in the DC-SQUID, which results in an effective contribution of the $X$-loop flux to the $Z$-loop~\cite{khezriAnnealpathCorrectionFlux2021}. Assuming negligible $X$-loop inductance, the $Z$-loop symmetry point of the tunable CSFQ is found to be (see e.g. Supplemental Information in Ref.~\cite{khezriAnnealpathCorrectionFlux2021})
\begin{align}
     \frac{\Phi_z^\t{sym}}{\Phi_{0}}&=0.5 + \frac{1}{2\pi}\arctan{\left[d\tan(\frac{\pi\Phi_x}{\Phi_0})\right]},\label{eq:Asymmetry}
\end{align}
where $d=(I_{cl}-I_{cr})/(I_{cl}+I_{cr})$ is the junction asymmetry, with $I_{cl(r)}$ being the critical current of the left(right) junction in the DC-SQUID. The symmetry point can be extracted by checking the point of reflection symmetry for each trace of the transmission measurement at each value of $\Phi_x$. Except for the point at $\Phi_x=0.1~\Phi_0$, which we attribute to anti-crossing between the resonator and the CSFQ higher energy levels and crosstalk calibration errors, the extracted symmetry points fits well to an asymmetry value of $d=0.069\pm 0.001$. To verify the effect of finite $X$-loop inductance, we use a full circuit model (see Supplementary Note 3) to numerically extract the symmetry points and compare with the analytical expression. As shown in Fig.~\ref{fig:Calibration}(b), using the same junction asymmetry $d$, the discrepancy between the numerically simulated and analytical $\Phi_z^\t{sym}$ grows larger as $\Phi_x$ increases, but always remains below $1.5~\text{m}\Phi_0$ for the range where the coherence measurements were performed, justifying the use of the simpler analytical expression for extracting the junction asymmetry from the data.

Once the crosstalk and junction asymmetry are calibrated, spectroscopy of the qubit is performed at a range of $\Phi_z$ and $\Phi_x$ biases, with the resonator SQUID biased at $\Phi_r=0$. A circuit model of the device is fit to the transition frequencies between ground, first and second excited states. The model includes the full capacitance matrix between islands, the loop inductances, and the Josephson junction critical currents, and is simulated using a package developed in Ref.~\cite{kerman_2020_efficientnumericalsimulation}. The fit has been presented in the supplemental material of a previous work~\cite{dai_dissipative_2022}, and the circuit parameters are reproduced in Supplementary Note 3 for completeness. 

Using the circuit model, the qubit persistent current $I_p$ and tunneling amplitude $\Delta$ versus the flux bias in the $X$-loop $\Phi_x$ can be obtained. The persistent current $I_p$ can be calculated using two different methods. In the first method, we rely on the operator matrix element of the circuit, which corresponds to the magnetic dipole moment of the CSFQ, given by
\begin{align}
I_{p,\t{c}}&=\bra{0_\t{c}}\frac{\partial H_\t{c}}{\partial \Phi_z}\ket{1_c},
\end{align}
where $\ket{0_\t{c}}(\ket{1_\t{c}})$ is the circuit ground (excited) state, and $H_\t{c}$ is the circuit Hamiltonian, all evaluated at the symmetry point $\Phi_z^\t{sym}$.

\begin{figure}
    \centering
    \includegraphics[width=3.4in]{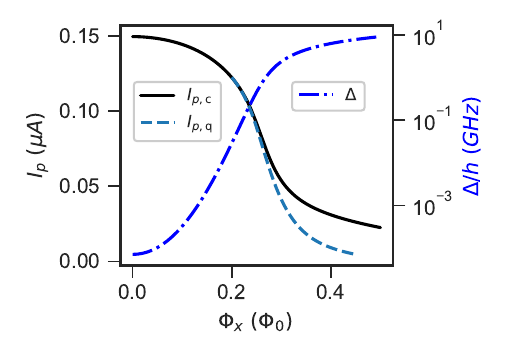}
    \caption{Simulated qubit persistent current and energy gap as a function of flux bias. Left axis, the simulated qubit persistent current $I_p$ versus $X$-loop external flux bias $\Phi_x$, obtained using the circuit operator (black) and fitting to two-state approximation (light blue dashed) (see subsection "Device characterization" for detail). Right axis, the simulated qubit tunneling amplitude $\Delta$ versus $X$-loop external flux bias $\Phi_x$.}
    \label{fig:IpDelta}
\end{figure}

In the second method, we find the ground-to-excited-states transition frequencies and fit to the qubit frequencies given by the two-state Hamiltonian Eq.~\ref{eq:TwoLevelHam}. The corresponding persistent current is denoted as $I_{p,\t{q}}$. In Fig.~\ref{fig:IpDelta} we show $I_{p,\t{c}}$ and $I_{p,\t{q}}$ versus $\Phi_x$. It can be seen that they agree with each other for small values of $\Phi_x$. At larger $\Phi_x$, $I_{p,\t{q}}$ becomes smaller than $I_{p,\t{c}}$. This can be seen as a result of the breakdown of the two-state approximation, as the lowest energy states of the circuit are no longer spanned by two well-defined persistent current states as $\Phi_x$ (or equivalently the tunneling amplitude $\Delta$) increases. This could lead to deviation from the transverse field Ising model typically used to describe a quantum annealer. For solving optimization problems with adiabatic evolution, such deviation likely has little impact, as long as the final problem Hamiltonian is accurately represented. However, for other applications such as quantum simulation or evolutions involving non-adiabatic transitions~\cite{munoz-bauza_2019_doubleslitproposalquantum, khezri_2022_customizedquantumannealing,fry-bouriaux_2021_locallysuppressedtransversefield}, the deviation of the CSFQ Hamiltonian from the simple spin model presents a complication that is worth being studied further.   

When compared with rf-SQUID flux qubits used in commercially available annealers, the CSFQ presented here has a persistent current that is at least an order of magnitude smaller, hence reducing its sensitivity to flux noise. The smaller $I_p$ also leads to a smaller interaction strength between qubits, which potentially limits the optimization problems that can be mapped to the annealer. We note that this could be compensated by galvanic coupling, using shared segments of loops~\cite{weber_2017_coherentcoupledqubits}, and even shared junction or junction arrays between the circuits that need to be coupled. 

The simulated tunneling amplitude $\Delta$, the qubit frequency at the symmetry point, is shown in Fig.~\ref{fig:IpDelta}. To implement a standard annealing experiment, the CSFQ is initialized at a large $\Delta$, around $\Phi_x\approx 0.5~\Phi_0$, and then the tunneling amplitude is gradually reduced following a specific annealing schedule. The CSFQ has a more gentle dependence of $\Delta$ on $\Phi_x$ as compared to the rf-SQUID qubit. Given the same resolution in the biasing sources, this $\Delta$ dependence allows finer control of the annealing schedule, and assists in future experimental demonstrations of novel annealing protocols, such as those proposed in Ref.~\cite{munoz-bauza_2019_doubleslitproposalquantum,khezri_2022_customizedquantumannealing}. 

\subsection*{Relaxation times}
In this subsection we present the measurements of the relaxation times $T_1$ of the CSFQ. We note that $T_1$ (and similarly $T_\phi$) coherence times are only well-defined when the system is weakly coupled to the environment, which is commonly the case in devices made for GMQC. In the context of annealing, the validity of the weak-coupling approximation depends on the relative strength of the qubit fields and the noise. In this work, we measure the coherence times for qubit frequency down to $\Delta/(2\pi)\approx 1~\t{GHz}$, below which thermally excited qubit population becomes significant. Although qubit state initialization at lower qubit frequencies is possible through high-fidelity single-shot readout~\cite{johnson_2012_heraldedstatepreparationa}, we do not pursue them in this work, as the available measurement range is enough to determine the strengths of the different noise sources, and for small enough $\Delta$, the weak-coupling limit eventually breaks down~\cite{dai_dissipative_2022}.

\begin{figure}
    \centering
    \includegraphics[]{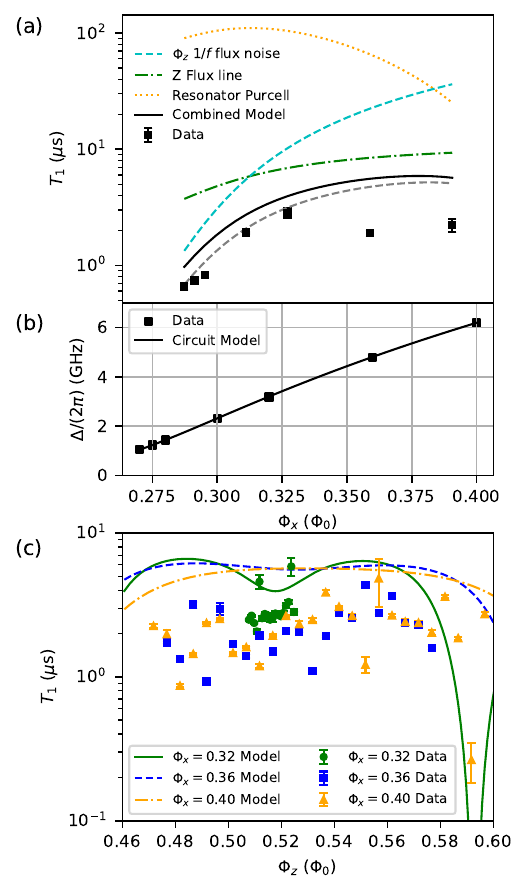}
    \caption{Relaxation time $T_{1}$ as a function of flux bias. (a) Measured (black square markers) and simulated $T_{1}$ values as a function of $\Phi_x$, at the $\Phi_z$ symmetry point. Each measured point is the result of averaging 30 repeated measurements, and error bars are the standard deviation. The simulated $T_1$ considered contributions from different sources, with the combined simulated $T_1$ shown in black. The grey dashed line shows (for reference only) the simulated $T_1$ decay rate if the intrinsic flux follows $1/f^{0.96}$ instead of $1/f$ spectrum. (b) Measured (black square markers) and simulated (lines) qubit frequency $\Delta$ as a function of $\Phi_x$, at the $\Phi_z$ symmetry point. (c) Measured (solid markers) and simulated (lines) $T_{1}$ as a function of $\Phi_z$ at $\Phi_x = 0.32~\Phi_0$ (green circles), $0.36~\Phi_0$ (blue squares), and $0.4~\Phi_0$ (orange triangles).  The simulated $T_1$ combines all noise sources shown in panel (a). There are no free parameters in the simulation; the flux noise amplitudes are obtained by fitting to the dephasing measurements for a fixed value of $\alpha$. See "Dephasing times" subsection in the text.}
    \label{fig:T1}
\end{figure}

The relaxation time $T_1$ was measured by initializing the qubit to the first excited state with a $\pi$ pulse and then performing readout after a variable delay time. The duration of the $\pi$ pulse (and $\pi/2$ pulse for the Ramsey measurements discussed in the "Dephasing times" subsection) were calibrated from Rabi oscillation measurements. While the specific pulse widths were bias-dependent, typical values are 10 ns for the $\pi$ pulses and 5 ns for the $\pi/2$ pulses. The relaxation time $T_1$ is measured both as a function of $\Phi_x$, at the $Z$-loop symmetry point $\Phi_z^\text{sym}$, as shown in Fig.~\ref{fig:T1}(a), and as a function of $\Phi_z$ near the symmetry point at three different $\Phi_x$, as shown in Fig.~\ref{fig:T1}(c). At each $\Phi_x$, the relaxation measurement (and dephasing measurement presented later) at the symmetry point is repeated 30 times, and the average value is reported. We see that as $\Phi_x$ varies from $0.27~\Phi_0$ to $0.4~\Phi_0$, corresponding to $\Delta$ changing from $1~\t{GHz}$ to $6.2~\t{GHz}$ (see Fig.~\ref{fig:T1}(b)), the $T_1$ at the $Z$-loop symmetry point increases at first, reaching a maximum at $\Phi_x=0.32~\Phi_0$, and then decreases. This is similar to other coherence studies on flux qubits and fluxoniums~\cite{yan_2016_fluxqubitrevisited,quintana_2017_observationclassicalquantumcrossover,sunCharacterizationLossMechanisms2023}, where the qubit coherence is limited by $1/f$ flux noise at lower frequency, and transitions to a different noise channel at frequencies higher than $\sim 1~\t{GHz}$. Furthermore, we also observe that for $\Phi_x=0.32,~0.36,~0.4~\Phi_0$, the measured relaxation times are nearly independent of the $Z$-bias $\Phi_z$ near the symmetry point, which also indicates that the $Z$-loop noise is not the dominant noise channel at higher values of $\Phi_x$. 

To better understand the relaxation data, we consider several relaxation channels and show their individual and combined calculated relaxation times in Fig.~\ref{fig:T1}, including intrinsic flux noise in the $Z$-loop, thermal noise from the $Z$-loop bias line and Purcell effect due to the readout resonator. We assume the intrinsic flux noise has a noise PSD of $2\pi A_{\Phi_z}/\omega$, with $\omega$ being the angular frequency and $A_{\Phi_z}$ the noise power at $\omega/(2\pi)=1~\text{Hz}$, set by fitting to the dephasing times (see "Dephasing times" subsection). The flux-bias line can be modelled as a 50-$\Omega$ impedance in parallel with a bias inductor that is coupled to the qubit loops. The noise temperature of the bias line can be approximately estimated based on the attenuations used along the signal line~\cite{krinnerEngineeringCryogenicSetups2019}. The relaxation time due to the Purcell effect is estimated using the relation for coupling to a single-mode resonator and is given by~\cite{houckControllingSpontaneousEmission2008}
\begin{align}
    T_1^\t{Purcell}&=\left[\frac{(\omega_r-\omega_{01})}{ g}\right]^2\kappa^{-1}.
\end{align}
Here $\omega_r$, $\omega_{01}$, and $g$ are the resonator, qubit transition frequencies and the linear coupling strength between the qubit and the resonator. The coupling strength $g$ is calculated using the circuit model parameters (see Supplementary Notes 3 and 4 for the qubit and resonator models), and is found to range from about $100~\t{MHz}$ to $200~\t{MHz}$ depending on the flux bias. The resonator decay rate $\kappa$ is estimated based on resonator linewidth measurements. At the readout point, the resonator frequency and decay rate are $\omega_r/(2\pi)=7.89~\text{GHz}$ and $\kappa/(2\pi)=1.9~\text{MHz}$. Besides the above noise channels, we also note that the intrinsic flux noise in the $X$-loop has negligible contribution to relaxation due to the comparatively small value of the X flux bias matrix elements (see discussion in Supplementary Note 5). The microwave port of the $X$-bias line, which became disconnected during the experiment, could have lead to additional relaxation due to thermal noise on the $X$-bias line. This is estimated to make substantial contribution to relaxation for $\Phi_x\gtrsim0.4~\Phi_0$, limiting $T_1$ to about $2~\mu\text{s}$. Details of the relaxation calculations are given in the Methods section. As shown in Fig.~\ref{fig:T1}, at each $X$-flux bias $\Phi_x$, the predicted relaxation time by combining the above known sources of noise comes within a factor of 2 of the highest $T_1$ measured at that value of $\Phi_x$. 

We discuss the potential causes for the disagreement between the experimental data and the model used for energy relaxation. First, at lower $\Phi_x$, the $1/f$ noise power could differ from the noise power extrapolated using the dephasing time fits based on the simple $1/f$ dependence. Indeed, measurements done on devices fabricated using similar process show that the intrinsic flux noise has $1/f^\alpha$ dependence, with $0.9\lesssim\alpha\lesssim1$~\cite{yan_2016_fluxqubitrevisited,braumuller_2020_characterizingoptimizingqubit}. To find a consistent set of $\alpha$ and flux noise amplitude values that explain both the relaxation and dephasing time data, we fit the flux noise amplitudes to the dephasing data with different values of $\alpha$, between 0.85 to 1.05 in steps of 0.01. Assuming there are stochastic noise channels such as two-level systems that the model does not account for, we only consider sets of $\alpha$ and flux noise amplitudes where the experimental $T_1$ values are strictly lower than the modeled ones. As indicated in Fig.~\ref{fig:T1}, we found if the dephasing times are fit instead with $\alpha=0.96$, the model predicts lower relaxation times than when choosing $\alpha=1$, in better agreement with the experimental data at lower $\Phi_x$. Second, at higher $\Phi_x$, or qubit frequency, we note that there is significant scatter in the measured $T_1$ data, especially at values of $\Phi_x>0.32~\Phi_0$. This can be seen from the $T_1$ versus $\Phi_z$ measurements shown in Fig.~\ref{fig:T1}(c). The scatter in the $T_1$ data could be due to coupling to microscopic two-level systems (TLS). Although the underlying physical nature of TLSs are still under active investigation~\cite{mullerUnderstandingTwolevelsystemsAmorphous2019}, resonant coupling to TLSs usually leads to variations in $T_1$ over qubit frequencies~\cite{barendsCoherentJosephsonQubit2013}, and sometimes time-dependent fluctuations~\cite{bejanin_2021_interactingdefectsgenerate}. This is also in line with recent measurement on fluxoniums, showing that TLSs indeed have a strong contribution to relaxation at high frequencies~\cite{sunCharacterizationLossMechanisms2023}. Besides TLSs, quasi-particles are known to cause fluctuations in $T_1$ over time~\cite{popCoherentSuppressionElectromagnetic2014}. To check the effect of quasi-particles, we compared $T_1$ fits using a simple exponential decay model and a model incorporating non-exponential decay due to quasi-particle tunneling during the $T_1$ measurement. The reduced chi-squared of the fits are almost the same in the two models, and therefore quasi-particle is likely not a dominant source of $T_1$ loss in our experiment.

Previous coherence measurement on flux qubits show that $T_1$ at higher qubit frequency is likely limited by intrinsic flux noise with an ohmic or superohmic noise spectrum~\cite{quintana_2017_observationclassicalquantumcrossover,sunCharacterizationLossMechanisms2023}, although it could be difficult to distinguish ohmic flux noise from charge noise~\cite{yan_2016_fluxqubitrevisited}. We found that including ohmic flux or charge noise in the model does not lead to substantially better agreement between the simulation and the measurement. This suggests that at higher qubit frequencies, thermal noise from the bias line is likely the main source of relaxation, along with TLSs which caused additional variations in the measured relaxation rates. The large bias line noise is due to the use of a large bandwidth low-pass filter at 12 GHz, opted to explore fast annealing and allow spectroscopic characterization through the bias lines. This could be mitigated by reducing the cutoff frequency to values of the order of 1 GHz, which retains the compatibility for most annealing protocols, while suppressing the bias line induced relaxation rate to below the intrinsic flux noise limit.
 
\subsection*{Dephasing times}\label{sec:Dephasing}
\begin{figure}
\centering
\includegraphics{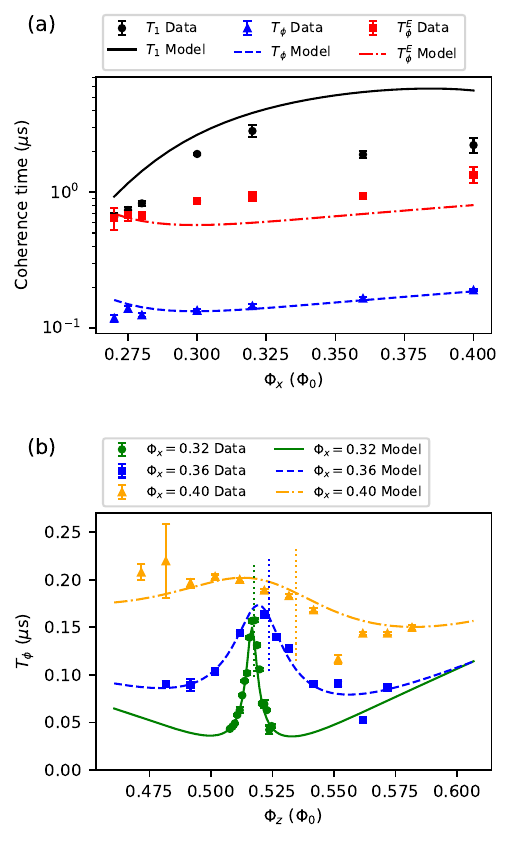}
\caption{Dephasing time as a function of flux bias. (a) Measured (solid markers) and simulated (lines) relaxation time $T_1$ (black circles), Ramsey $T_\phi$ (blue triangles) and spin-echo $T_\phi^E$ (red squares) dephasing times, as a function of $\Phi_x$, with $\Phi_z$ set at the symmetry point. Each measured point is the result of averaging 30 repeated measurements, and error bars are the standard deviation. (b) Measured (solid markers) and simulated values (lines) of $T_\phi$ as a function of $\Phi_z$ at $\Phi_x = 0.32~\Phi_0$ (green circles), $0.36~\Phi_0$ (blue squares), and $0.4~\Phi_0$ (orange triangles). The vertical dashed lines indicate the position of the $Z$-loop symmetry point at each value of $\Phi_x$.}
\label{fig:T2}
\end{figure}

We next discuss the dephasing measurements, which were also performed at a range of $X$ and $Z$ flux biases. $T_\phi$ times were measured using a Ramsey protocol (see for example Ref.~\cite{ithier_2005_decoherencesuperconductingquantum}) which consists of initializing the qubit with a $\pi$/2 pulse (detuned approximately 10 MHz from the qubit transition frequency) and then applying a second pulse after a variable delay time just before performing readout. As shown in Fig.~\ref{fig:T2}(a), at the $Z$ bias symmetry point, the Ramsey pure dephasing time $T_\phi$ varies from $\sim200~\t{ns}$ to $\sim 100~\t{ns}$ as $\Phi_x$ is reduced from $0.4~\Phi_0$ to $0.27~\Phi_0$. The spin-echo dephasing time $T_\phi^E$, measured with a sequence that has an additional $\pi$ pulse in between the two $\pi/2$ pulses~\cite{ithier_2005_decoherencesuperconductingquantum}, shows an improvement of about a factor of $5$. This improvement, and the fact that all the measured coherence decays have been well-fitted with a Gaussian envelope, are consistent with dephasing dominated by $1/f$ noise. When examining the $\Phi_z$ dependence of $T_\phi$ as shown in Fig.~\ref{fig:T2}(b), it is found that the maximum dephasing time occurs slightly to the left of the $Z$ bias symmetry point $\Phi_z^\t{sym}$, which stands in contrast to the non-tunable flux qubit. 

To model dephasing (see Methods section), we assume that $1/f$ flux noise in the $Z$- and $X$-loops are the only sources of dephasing, with noise power spectral densities $S_{\Phi_z}=2\pi A_{\Phi_z}/\omega$ and $S_{\Phi_x}=2\pi A_{\Phi_x}/\omega$ respectively. In addition, we assume the flux noise in the $Z$- and $X$-loops have a correlated noise PSD $C_{\Phi_z\Phi_x}$, represented by the dimensionless coefficient, $c_{zx}=C_{\Phi_z\Phi_x}/\sqrt{S_{\Phi_z}S_{\Phi_x}}$. The frequency sensitivity to flux noise is determined directly from the circuit model, without the two-state approximation (see Supplementary Note 5). The noise powers and correlation coefficients are fitted to the $\Phi_z$ dependence of the dephasing times $T_\phi$. As shown in Fig.~\ref{fig:T2}(b), the model describes the flux dependence of the dephasing times well, indicating that the dephasing times are indeed limited by noise sources that couple to the flux biases of the circuits. The noise powers that fit the measured $T_\phi$ best are $\sqrt{A_{\Phi_z}}=13.4~\mu\Phi_0/\sqrt{\text{Hz}}$, $\sqrt{A_{\Phi_x}}=7.4~\mu\Phi_0/\sqrt{\text{Hz}}$. These numbers are consistent with previous devices fabricated using a similar process~\cite{weber_2017_coherentcoupledqubits}, considering that flux noise power scales as length over width of the loop wires~\cite{lanting_2009_geometricaldependencelowfrequency,braumuller_2020_characterizingoptimizingqubit}. 

When applying the same model to the $\Phi_x$ dependence of the dephasing times $T_\phi$ and $T_\phi^E$, as shown in Fig.~\ref{fig:T2}(a), the model agrees relatively well with the Ramsey dephasing times $T_\phi$ and underestimates the dephasing times measured by spin-echo. As the spin-echo sequence is less sensitive to low-frequency noise ($\omega\lesssim T_2^E\sim 1~\mathrm{MHz}$, the disagreement between the measured and calculated spin-echo dephasing times could indicate that the flux noise is more concentrated in the low-frequency range than what the $1/f$ spectrum predicts. One possible explanation could be excess electronics noise, though further experiments are required to verify this. 

The best fit of the dephasing times indicates a correlation coefficient of $c_{zx}=0.49$ (see Methods section for definition). Flux noise correlation in tunable flux qubits has been measured previously~\cite{gustavssonNoiseCorrelationsFlux2011,sunCharacterizationLossMechanisms2023}, and it has been pointed out that positive (negative) correlation shifts the maximum dephasing time to the left (right) of the symmetry point. Assuming flux noise arises from uniformly distributed environmental spins on the metal surface of the superconducting loop, Ref.~\cite{gustavssonNoiseCorrelationsFlux2011} shows that the correlation could be understood in terms of spins that occupy the shared arm between the $Z$- and $X$-loops. However, given the symmetrized $X$-loop design used here, the expected correlation using this simple model for flux noise is zero, and hence does not explain the measured correlation. 

Junction asymmetry could also lead to an offset between the maximum dephasing time and the qubit symmetry point. To see this, one could define another effective $Z$-flux bias $\Phi_{\tilde{z}}$, such that the symmetry point in $\Phi_{\tilde{z}}$ is independent of $\Phi_x$. This means
\begin{align}
    \Phi_{\tilde{z}}=\Phi_{z}-F(\Phi_x),
\end{align}
where $F$ is given by Eq.~\ref{eq:Asymmetry} when the $X$-loop inductance is negligible. Then it can be shown that the noise correlation coefficient between $\Phi_{\tilde{z}}$ and $\Phi_x$ is given by (see Supplementary Note 1)
\begin{align}
    c_{\tilde{z}x}=c_{zx}-\frac{dF}{d\Phi_x}\frac{S_{\Phi_x}}{\sqrt{S_{\Phi_z}S_{\Phi_x}}}.
\end{align}
where $c_{zx}$ is the dimensionless correlation between $\Phi_z$ and $\Phi_x$ flux noise. Therefore, for negligible noise correlation $c_{zx}$ between the $Z$- and $X$- loops, as would be the case if noise purely comes from uniformly distributed spins on the metal surface, the correlation coeffcient $c_{\tilde{z}x}$, between the effective $Z$-loop and $X$-loop would be negative given a positive junction asymmetry $d$. In other words, the effect of junction asymmetry in the device is to shift the point of maximum dephasing time to the right of the symmetry point, which is the opposite of what is observed experimentally. We note that this apparent offset should not be confused with the fitted correlation $c_{zx}$, as the fitting uses numerically extracted frequency sensitivity to $\Phi_z$ and $\Phi_x$, which implicitly contains information about the asymmetry. 

To further investigate the source of the apparent noise correlation between the $Z$- and $X$-loops, we considered additional sources of dephasing, including bias line thermal noise, voltage noise of the bias source, photon shot noise, second-order coupling to flux noise, charge noise and junction critical-current noise (see Supplementary Notes 6 and 7). We find that using previously reported values of $1/f$ noise of the critical current of the junctions~\cite{yan_2012_spectroscopylowfrequencynoise}, its contribution to dephasing is about an order of magnitude smaller compared to dephasing due to first-order coupling to flux noise. Furthermore, junction critical-current noise leads to a maximum dephasing time to the left of $\Phi_z^\t{sym}$. This means the junction critical current noise could contribute to the apparent positive correlation between the two flux bias noise sources, though is likely not the dominant source. In addition, we note that in contrast to previous studies, where the flux noise correlation could mostly be explained by environmental spins local to the metal surface of the superconducting loop~\cite{gustavssonNoiseCorrelationsFlux2011,sunCharacterizationLossMechanisms2023}, our device has a flip-chip architecture, involving a ground plane on the interposer chip facing the loops. This can be seen in Fig.~\ref{fig:CircuitChip}(b), while the interposer ground plane facing the main part of the qubit $Z$-loop are designed with cutouts to allow large geometric mutual inductance, the region facing the qubit $X$-loop does not have such cutouts. This design difference points to the possibility of screening effects and environmental spins on the interposer ground plane as a source of noise correlation. Qualitatively it could be understood as follows: Persistent currents in the flux qubit loops induce screening currents on the interposer chip. Fluctuating environmental spins on the interposer chip could interact with the screening currents, which induces fluctuations in the magnetic field experienced by the qubit loops due to the screening current. Since the distance between the interposer chip and the qubit chip is only about $3~\mu$m, comparable to the width of the qubit loop wires, we expect the magnitude of the screening current to be on the same order as the persistent current. Furthermore, due to the different geometries on the qubit chip and the interposer chip, the screening currents will likely lead to different noise correlations as compared to the persistent current on the qubit loops. Therefore, the overall noise correlation observed in the experiment could be different from the one we expect if flux noise only comes from noise sources local to the flux qubit loops. The precise effects of the environmental spins on the interposer chip will be investigated in future studies. 

We also observed over multiple cooldowns of the dilution fridge and changes to the setup that the choice of bias source could have an impact on the coherence times as well. Experiments performed during earlier cooldowns used AWGs as the DC bias sources for the device. After switching from using AWGs to a lower noise DC bias source for the $X$-bias, we found the measured Ramsey pure dephasing time at $\Phi_x=0.32~\Phi_0$ improved from $\sim 80$ ns initially to $\sim150$ ns after the switch. In Supplementary Note 6, we show that by assuming the AWG has a combined $1/f$ and white noise spectrum at low frequency, it alone could lead to a dephasing time of $T_\phi\sim 350$ ns. This is roughly consistent with the reduction in Ramsey dephasing time when the AWG is used, given the crude model of the AWG noise. As the DC bias source does not support fast annealing, strategies need to be developed to mitigate the noise from the AWG. This could be done by either using a cryogenic bias-T to combine the DC and fast signal, allowing a smaller coupling strength of the more noisy signal, or applying heavier filtering to the fast signal and correct for distortions in-situ~\cite{krinner_2022_realizingrepeatedquantuma}. 

\subsection*{Noise in annealing parameters}\label{sec:AnnealingNoise}
\begin{figure}
    \centering
    \includegraphics[width=3.4in]{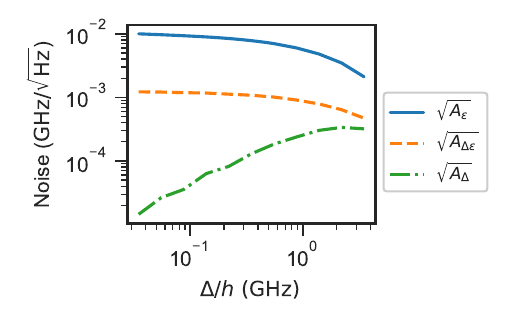}
    \caption{Noise in annealing parameters. 1/f noise amplitudes $A_\epsilon$ (blue solid line) and $A_\Delta$ (orange dotted line) at 1 Hz for the annealing parameters $\epsilon$ (longitudinal field) and $\Delta$ (transverse field) as a function of $\Delta$, calculated from the measured 1/f flux noise in the two control loops. The green dash-dotted curve shows the correlated noise amplitude $A_{\Delta\epsilon}$.}
    \label{fig:AnnealingNoise}
\end{figure}

It is useful to put the noise measured here in the context of annealing applications. For a ``single-qubit anneal"~\cite{khezriAnnealpathCorrectionFlux2021,morrellSignaturesOpenNoisy2022}, the qubit Hamiltonian starts with a large value of $\Delta$, at which the qubit is initialized in the ground state, and the qubit Hamiltonian is gradually changed to a target Hamiltonian where $\Delta$ is close to zero. The only parameter defining the target Hamiltonian is the qubit longitudinal field

\begin{align}
    \epsilon = \frac{2I_p(\Phi_z-\Phi_z^\t{sym})}{\hbar}.
\end{align}
Noise in the control fluxes gives rise to noise in the annealing parameters $\epsilon, \Delta$, with the same frequency dependence. We focus on the intrinsic $1/f$ flux noise, assuming noise arising from wiring and electronics can be minimized through careful engineering. Using the two-state approximation Hamiltonian Eq.~\ref{eq:TwoLevelHam}, we find the individual and correlated $1/f$ noise in $\epsilon, \Delta$ to be
\begin{align}
    A_\epsilon =& \left(\frac{\partial\epsilon}{\partial\Phi_z}\right)^2A_{\Phi_z}+\left(\frac{\partial\epsilon}{\partial\Phi_x}\right)^2A_{\Phi_x}\nonumber\\
    &+2\frac{\partial \epsilon}{\partial \Phi_z}\frac{\partial \epsilon}{\partial \Phi_x}A_{\Phi_z\Phi_x},\\
    A_\Delta =&\left(\frac{\partial \Delta}{\partial \Phi_x}\right)^2A_{\Phi_x},\\
    A_{\Delta\epsilon}=&\frac{\partial \epsilon}{\partial \Phi_z}\frac{\partial \Delta}{\partial \Phi_x}A_{\Phi_z\Phi_x}+\frac{\partial \epsilon}{\partial \Phi_x}\frac{\partial \Delta}{\partial \Phi_x}A_{\Phi_x}^2,
\end{align}
where $A_{\Phi_z\Phi_x}=c_{zx}\sqrt{A_{\Phi_z}A_{\Phi_x}}$ is the $1/f$ power of the flux noise correlation. In Fig.~\ref{fig:AnnealingNoise} we show these three quantities versus the transverse field $\Delta$, for the target longitudinal field $\epsilon=0$, noting these quantities are weakly dependent on $\epsilon$. It can be seen that the $\epsilon$ noise, $A_\epsilon$ is always the dominant noise factor. The correlated noise $A_{\Delta\epsilon}$ is mainly due to the measured flux noise correlation, and is about an order of magnitude lower than $A_\epsilon$. The noise in $A_\Delta$ is comparable to $A_{\Delta\epsilon}$ for large $\Delta$, but diminishes quickly with decreasing $\Delta$. The role of the relative strength of the noises will be explored in future work. One interesting direction is to extend the previous analysis of the effect of correlated noise on single qubit Landau-Zener tunneling~\cite{saito_2007_dissipativelandauzenertransitions}, to the general case of multi-qubit annealing.

\section*{Conclusions}\label{sec:Conclusion}
We have presented a detailed characterization of coherence in a CSFQ design relevant for quantum annealing. We measured the $T_1$ relaxation and $T_\phi$ dephasing times and modelled them considering all relevant noise sources. We find that the $T_1$ values are influenced primarily by $1/f$ flux noise with additional contributions from bias line thermal noise and TLSs at higher frequencies. The measured dephasing time $T_\phi$ is dominated by $1/f$ flux noise, with a possible contribution from $1/f$ noise in the junction critical current. We point out that for annealing applications, the external flux biases can introduce significant relaxation and dephasing as compared to the intrinsic noise of the qubit. This suggests that future annealing experiments that aim to combine high-coherence and high control flexibility need to carefully evaluate the trade-off between the added noise and the improved control bandwidth. 

Additionally, we observe a positive correlation between the $Z$ and $X$ flux noise, as manifested in the maximum dephasing time versus $\Phi_z$ occurring to the left of the $Z$ bias symmetry point. This correlation may arise due to environmental spins on the interposer ground plane, and junction critical-current noise. Future experiments that are able to distinguish the different sources of $1/f$ noise are needed to quantitatively determine the origin of this noise correlation. 

Our work provides a detailed characterization of noise sources of a flux qubit, relevant for future analysis of coherent annealers based on CSFQs. Besides quantum annealing, the work presented here is also relevant for flux qubit devices made for GMQC, such as fluxonium qubits. Future work will be directed at extending the coherence characterization to the strong coupling limit and multi-qubit systems, as well as studying coherence in a dynamic setting, which will lead to a deeper understanding of the role of coherence in quantum annealing. 

\begin{acknowledgments}
We thank the members of the Quantum Enhanced Optimization (QEO)/Quantum Annealing Feasibility Study (QAFS) collaboration for various contributions that impacted this research. In particular, we thank A. J. Kerman for guidance on circuit simulations, S. Novikov for assisting the chip design, and K. Zick for coordinating the experimental efforts in the collaboration. We gratefully acknowledge the MIT Lincoln Laboratory design, fabrication, packaging, and testing personnel for valuable technical assistance. The research is based upon work supported by the Office of the Director of National Intelligence (ODNI), Intelligence Advanced Research Projects Activity (IARPA) and the Defense Advanced Research Projects Agency (DARPA), via the U.S. Army Research Office contract W911NF-17-C-0050. The views and conclusions contained herein are those of the authors and should not be interpreted as necessarily representing the official policies or endorsements, either expressed or implied, of the ODNI, IARPA, DARPA, or the U.S. Government. The U.S. Government is authorized to reproduce and distribute reprints for Governmental purposes notwithstanding any copyright annotation thereon.
\end{acknowledgments}

\section*{Competing Interests Statement}
The authors declare no competing interests.

\section*{Author Contributions Statement} 
R.T. performed the experiments. R.T. X.D., and M.A.Y. performed the data analysis. X.D. performed the numerical simulations. D.M., M.A.Y., A.J. Martinez, and Y.T. designed the device. J.G. J.A.G. and X.D. performed earlier versions of the experiments on a different device and setup. D.M.T., J.A.G., S.D., and J.B. contributed to the development of the experiment infrastructure.
R.D., D.K.K., A.J.Melville, B.M.N., and J.L.Y. developed the fabrication process and fabricated the device. C.H., K.S., and S.J.W. contributed to the fridge and electronics operation. W.O. supervised the QEO/QAFS effort from Lincoln lab, K.M.Z. led the coordination of the QEO/QAFS experimental effort, D.L. led the QEO/QAFS program and A.L. proposed and supervised this work. All authors were involved in the discussion of experiments and data analysis. X.D., R.T., and A.L. wrote the paper with feedback from all authors.

\section*{Data Availability}
The data presented in this study have been deposited at Ref.~\cite{figshare}. 

\section*{Code Availability}
The code for analyzing the decoherence data is available at Ref.~\cite{dai_2024_highrizercsfq_decoherence}.

\section*{Methods}
\subsection*{Noise power spectral density and correlation}\label{sec:NoisePSDAndDecoherence}
In this work, we define the noise power spectral density (PSD) of a random variable $\delta \lambda$ as the Fourier transform of its autocorrelation function, 
\begin{align}
    S_{\lambda}(\omega) = \int^\infty_{-\infty}\text{d}\tau e^{i\omega \tau}\langle \delta\lambda(\tau) \delta\lambda(0)\rangle.
\end{align}
If the noise arises from a quantum bath, the noise PSD is asymmetric in positive and negative frequencies. However, our noise measurement is only sensitive to the symmetrized noise spectrum, which we define as 
\begin{align}
    S^+_{\lambda}(\omega)&=\frac{1}{2}\int^\infty_{-\infty}\text{d}\tau e^{i\omega \tau}\langle \delta\lambda(\tau) \delta\lambda(0)+\delta\lambda(0) \delta\lambda(\tau)\rangle\\
    &=\frac{1}{2}\l(S_{\lambda}(\omega)+S_{\lambda}(-\omega)\r).
\end{align}
The symmetrized correlation between two random variables $\delta \kappa$ and $\delta \lambda$ is given by the cross power spectral density
\begin{align}
    C_{\lambda \kappa}(\omega) &= \frac{1}{2}\int^\infty_{-\infty}\text{d}t e^{i\omega t}\left(\langle \delta\lambda(t) \delta\kappa(0) + \delta\lambda(0)\delta\kappa(t) \rangle\right)\\
    &\equiv c_{\lambda\kappa}\sqrt{S^+_{\lambda}(\omega)S^+_{\kappa}(\omega)},
\end{align}
where $c_{\lambda\kappa}$ is a dimensionless number describing the relative correlation between the two noise variables. 

\subsection*{Relaxation time calculation}\label{app:Relaxation}
We consider the qubit relaxation time $T_1$ in general to be given by 
\begin{align}
    \frac{1}{T_1}&=\frac{1}{T_1^\t{Purcell}}+\frac{1}{T_1^\t{z1f}}+\frac{1}{T_1^\t{x1f}}+\frac{1}{T_1^\t{zb}}+\frac{1}{T_1^\t{xb}}\\
    &+\frac{1}{T_1^\t{QOhmic}}+\frac{1}{T_1^\t{zOhmic}}+\frac{1}{T_1^\t{xOhmic}},
\end{align}
where $T_1^\t{Purcell}$ is the Purcell effect decay rate to the resonator, $T_1^\t{z(x)1f}$ are the relaxation times due to the intrinsic $1/f$ flux noise in the $Z(X)$ loops, $T_1^\t{z(x)b}$ are the relaxation times due to Johnson-Nyquist noise in the $Z(X)$ bias lines, and $T_1^\t{Q(z,x)Ohmic}$ are additional ohmic noises that couple to the qubit charge(flux) degrees of freedom~\cite{lanting_2011_probinghighfrequencynoise,yan_2016_fluxqubitrevisited,quintana_2017_observationclassicalquantumcrossover}. 

The relaxation time due to Purcell loss is estimated by~\cite{houckControllingSpontaneousEmission2008}
\begin{align}
    T_1^\t{Purcell}&=\frac{g^2}{(\omega_r-\omega_{01})^2}\kappa,
\end{align}
where $g$ is the exchange interaction strength between the qubit and the resonator, $\omega_r$ is the resonator frequency and $\kappa$ is the resonator decay rate. 

For the rest of the noise channels, we assume the noise is weak and the decay rate from the Bloch-Redfield theory (or equivalently Fermi's Golden rule) is given by~\cite{ithier_2005_decoherencesuperconductingquantum},
\begin{align}
    \frac{1}{T_1^\lambda}&=|\bra{1}\frac{\partial H_q}{\partial \lambda}\ket{0}|^2\left[S_\lambda(\omega_{01})+S_\lambda(-\omega_{01})\right],
\end{align}
where $\partial H_q/\partial \lambda$ is understood as the qubit operator that is coupled to the noise, and $S_\lambda(\omega_{01})$ is the noise PSD of $\lambda$ at qubit frequency $\omega_{01}$. For the intrinsic flux noise, we assume they are of the form $2\pi A_\lambda/\omega$, with the same noise amplitude as found in the pure dephasing rate fits. For the Johnson-Nyquist noise, we consider the bias line as an inductor with inductance $L_b$ shunted by an impedance of $Z_0=50~\Omega$. The current noise is given by
\begin{align}
    S_I(\omega) &= \frac{\hbar\omega\text{Re}[Z_b](1+\coth{\frac{\hbar\omega}{2k_BT_N}})}{\omega^2 L_b^2},\\
    \text{where}~&\frac{1}{Z_b}=\frac{1}{Z_0}+\frac{1}{i\omega L_b},
\end{align}
and $L_b=25~\text{pH}$ based on electromagnetic simulation. To compute the noise temperature $T_N$, we assume that each attenuator along the signal delivery chain acts as a beam-splitter~\cite{krinnerEngineeringCryogenicSetups2019}, so that the thermal noise photon number at stage $i$ is given as
\begin{align}
    n^{i}_N=A^{i}n_N^{i-1}+(1-A^{i})n_{\text{BE}}(T^i,\omega),
\end{align}
where $T^i$ and $A^i$ are the temperature of the fridge plate and the corresponding attenuation attached to that plate. The Bose-Einstein photon occupation number is given by $n_\text{BE}(T,\omega)=1/[\exp(\hbar\omega/k_BT)-1]$.
In particular for the $Z$-bias, we used $20$, $10$, $10$ dB attenuations at the $4~\t{K}$, $0.5~\t{K}$ and $10~\t{mK}$ stages of the dilution fridge. This attenuation scheme leads to an effective noise temperature of $170~\t{mK}$ at $\omega/(2\pi)=3~\text{GHz}$, which is relatively high compared to the base temperature of the fridge.

The additional $Q(z,x)$ ohmic noise can be given as
\begin{align}
    S_\text{Ohmic}=B\omega|\omega|^{\gamma-1}\left(1+\coth{\frac{\hbar\omega}{2k_BT}}\right),
\end{align}
with $\gamma=1$ and the different noise coefficients $B^Q, B^z, B^x$ corresponding to charge noise, $\Phi_z$ flux noise and $\Phi_x$ flux noise. Previous experiments on flux qubits have found that the dominant relaxation channel at higher qubit frequencies ($\gtrsim 1~\text{GHz}$) could be either ohmic charge or (super)ohmic flux noise~\cite{yan_2016_fluxqubitrevisited,quintana_2017_observationclassicalquantumcrossover,sunCharacterizationLossMechanisms2023}. In our experiments, we found that including the ohmic flux and charge noise do not substantially improve the agreement between the model and the simulation, due to the relatively significant scatter in our $T_1$ data.

\subsection*{Pure dephasing time Calculation}\label{sec:GeneralDecoherence}
Dephasing is caused by low-frequency fluctuations in the qubit energy splitting $\omega_{01}$ due to various noise sources. The noise spectrum of the qubit transition frequency $\omega_{01}$ is given by
\begin{align}
    S_{\omega_{01}}^+ & = \sum_\lambda \left(\frac{\partial \omega_{01}}{\partial\lambda}\right)^2S_{\lambda}^+ + 2\sum_{\lambda \neq \kappa}  \frac{\partial \omega_{01}}{\partial\lambda}\frac{\partial \omega_{01}}{\partial\kappa}C_{\lambda \kappa},\label{eqn:EnergyNoisePSD}
\end{align}
where the summation runs over the noise sources under consideration. For classical noise, dephasing can be described by the decay $\langle\exp{\left[-\chi(\tau)\right]}\rangle$~\cite{bylander_2011_noisespectroscopydynamical}, where $\chi$ is given by 
\begin{align}
    \chi(\tau)=\frac{\tau^2 }{2\pi} \int^{\infty}_{\omega_\text{low}} \text{d}\omega S_{\omega_{01}}^+(\omega) g_{N}(\omega\tau)\label{eqn:Dephasing}.
\end{align}
Here $g_N(\omega\tau)$ is the filter function for the specific coherence measurement, and $\omega_\text{low}$ is a low-frequency cutoff determined by the total experiment time. We have $N=0$ for Ramsey and $N=1$ for spin-echo measurement. Their respective filter functions are 
\begin{align}
    g_0 &= \text{sinc}^2\left(\frac{\omega \tau}{2}\right)~\text{and}\\
    g_1 &= \text{sinc}^2\left(\frac{\omega \tau}{4}\right)\text{sin}^2\left(\frac{\omega \tau}{4}\right).
\end{align}
In general the decay of $\langle\exp{\l[-\chi(\tau)\r]}\rangle$ is not exponential. We associate the pure dephasing time $T_\phi$ as the $1/e$ decay time, where $\langle\exp{\l[-\chi(\tau=T_\phi)\r]}\rangle=1/e$.

The intrinsic flux noise usually has a noise PSD of the form  
\begin{align}
    S_\lambda^+(\omega)=A_\lambda\l(\frac{2\pi}{|\omega|}\r)^\alpha\label{eqn:1/fNoise},
\end{align}
for $\lambda\in[\Phi_z,\Phi_x]$, where $\alpha\sim1$. If dephasing is dominated by flux noise, we can assume $S^+_{\omega_{01}}$ has the same frequency dependence, with
\begin{align}
    S_{\omega_{01}}^+&=A_{\omega_{01}}\l(\frac{2\pi}{|\omega|}\r)^\alpha.
\end{align}
In this case, the integral in Eq.~(\ref{eqn:Dephasing}) can be simplified, which leads to a dephasing time of
\begin{align}
    \frac{1}{T_\phi} &=\l(A_{\omega_{01}}\eta_N\r)^{1/(1+\alpha)}.\label{eqn:DephasingFitExpr}
\end{align}
Following Ref.~\cite{weber_2017_coherentcoupledqubits}, for Ramsey 
 and spin-echo measurements, the factors $\eta_0, \eta_1$ can be determined numerically by
\begin{align}
    \eta_0 &=(2\pi)^{\alpha-1}\int_{\omega_{\t{low}}t}^\infty \frac{dz}{z^\alpha}\l(\frac{\sin(z/2)}{z/2}\r)^2,\\
    \eta_1 &=(2\pi)^{\alpha-1}\int_{0}^\infty \frac{dz}{z^\alpha}\l(\frac{\sin(z/4)}{z/4}\r)^2\sin^2(z/4),
\end{align}
where $t$ is the typical free evolution time in a single Ramsey sequence. 

To fit the measured Ramsey dephasing times, Eqn.~\ref{eqn:DephasingFitExpr} is applied. The noise power $A_{\omega_{01}}$ is related to the flux noise powers via 
\begin{align}
    A_{\omega_{01}}=&\l(\frac{\partial\omega_{01}}{\partial \Phi_z}\r)^2A_{\Phi_z}+\l(\frac{\partial\omega_{01}}{\partial \Phi_x}\r)^2A_{\Phi_x}\\
    &+2\frac{\partial\omega_{01}}{\partial \Phi_z}\frac{\partial\omega_{01}}{\partial \Phi_x}c_{zx}\sqrt{A_{\Phi_z}A_{\Phi_x}},
    \end{align}
with $\partial \omega_{01}/\partial \Phi_{z(x)}$ extracted numerically from the circuit model. The factor $\eta_0$ is found by using $\omega_\t{low}=10\t{Hz}\times2\pi$ and $t=100\t{ns}$, and the noise exponent is assumed to be $\alpha=1$.

\section*{References}
\bibliography{references}

\end{document}